\documentclass{article}
\usepackage[utf8]{inputenc}

\usepackage{graphicx}
\usepackage{bm}
\usepackage{amsmath}
\usepackage{amssymb}
\usepackage{amsfonts}
\usepackage{amsthm}
\usepackage{mathtools}
\usepackage{hyperref}
\usepackage[left=1in,right=1in,top=1in,bottom=1in]{geometry}
\usepackage{dsfont}
\usepackage{comment}
\usepackage{cleveref}
\usepackage[dvipsnames]{xcolor}
\usepackage{authblk}

\hypersetup{colorlinks=true,urlcolor=[rgb]{0,0,0.5},citecolor=[rgb]{0.5,0,0},linkcolor=[rgb]{0,0,0.4}}

\newcommand{\SvecSq}{\vec{S}^2}
\newcommand{\SvecSqMax}{\vec{S}^2_{\mathrm{max}}} 
\newcommand{\ES}{\mathcal{E}}
\newcommand{\Id}{I}
\newcommand{\x}{\mathrm{x}}
\newcommand{\y}{\mathrm{y}}
\newcommand{\z}{\mathrm{z}}

\newcommand{\Psing}[1]{P_{\mathrm{sing}}^{(#1)}}

\newcommand{\sing}{\mathrm{sing}} 
\newcommand{\Ntot}{N_{\mathrm{tot}}} 
\newcommand{\n}{i}
\newcommand{\m}{k}
\newcommand{\elle}{\ell}
\newcommand{\w}{w}
\newcommand{\en}{n}

\newtheorem{lemma}{Lemma}

\title{A spin-energy operator inequality for Heisenberg-coupled qubits}
\author[1]{Daniel Ranard}
\author[2]{C. Jess Riedel}
\affil[1]{\footnotesize{Center for Theoretical Physics, Massachusetts Institute of Technology, Cambridge, MA 02139}}
\affil[2]{\footnotesize{Physics \& Informatics Laboratories, NTT Research Inc., 940 Stewart Drive, Sunnyvale, CA 94085}}
\date{\today}

\begin{document}

\maketitle

\begin{abstract}
We slightly strengthen an operator inequality identified by Correggi et al.\ that lower bounds the energy of a Heisenberg-coupled graph of $s=1/2$ spins using the total spin. In particular, $\Delta H \ge C \Delta\SvecSq$ for a graph-dependent constant $C$, where $\Delta H$ is the energy above the ground state and $\Delta\SvecSq$ is the amount by which the square of the total spin $\vec{S} = \sum_i \vec{\sigma}_i/2$ falls below its maximum possible value. We obtain explicit constants in the special case of a cubic lattice. We briefly discuss the interpretation of this bound in terms of low-energy, approximately non-interacting magnons in spin wave theory and contrast it with another inequality found by B\"arwinkel et al.
\end{abstract}

Consider a Hamiltonian defined on an arbitrary connected graph\footnote{We assume the graph does not have loops (i.e., edges that connect a vertex to itself), which would just change the Hamiltonian by a constant since $\vec{\sigma}_\n \cdot  \vec{\sigma}_\n = 3\Id$.} over a set of $\Ntot$ qubits where each pair of connected qubits experience the same Heisenberg coupling $J$:
\begin{align}
	H = -J \sum_{(\n,\m)\in \ES} \vec{\sigma}_\n \cdot  \vec{\sigma}_\m,
\end{align}
where the Pauli spin vector for the $\n$-th qubit is $\vec{\sigma}_\n =(\sigma^\x_\n, \sigma^\y_\n, \sigma^\z_\n)$.
Here the sum is over the edge set $\ES$ of all pairs $(\n,\m)$ of spins connected by edges on the graph.  Define $E_* := -J |\ES|$ where $|\ES|$ is the number of edges in the graph; this is the ground-state energy for the ferromagnetic case  $J>0$. 
Subtracting $E_*$ from the Hamiltonian we get
\begin{align}
	\Delta H :=& H - E_* 
	=  -J \sum_{(\n,\m)\in \ES} \left(\vec{\sigma}_\n \cdot  \vec{\sigma}_\m - \Id \right)
	= 4J \sum_{(\n,\m)\in \ES} \Psing{\n,\m}
\end{align}
where we have used $\vec{\sigma}_\n \cdot  \vec{\sigma}_\m = I - 4\Psing{\n,\m}$ for $i \neq k$.  Here,
$\Psing{\n,\m}:= |\sing\rangle_{\n,\m}\langle\sing|$ 
is the projector onto the singlet state $|\sing\rangle_{\n,\m}:=\left(| \uparrow\rangle_\n | \downarrow\rangle_\m - | \downarrow\rangle_\n | \uparrow\rangle_\m \right)/\sqrt{2}$ for qubits $\n$ and $\m$.  
In other words, the operator $\Delta H/4J$ is just the sum of all singlet projectors for spin pairs connected by Heisenberg interactions.  

Given the total spin operator $\vec{S} = \sum_{\n} \vec{\sigma}_{\n}/2$, let 
\begin{align}
    \label{eq:simple-start-periodic}
    \Delta\SvecSq :=& \SvecSqMax-\SvecSq 
    =\frac{\Ntot}{2}\left(\frac{\Ntot}{2}+1\right) - \frac{1}{4}\sum_{\n} \sum_{\m}  \vec{\sigma}_{\n}\cdot \vec{\sigma}_{\m}
    = \sum_{\n} \sum_{\m \neq \n}  \Psing{\n,\m} 
\end{align}
be the difference between the squared total spin and its maximum eigenvalue $ \SvecSqMax := (\Ntot/2)(\Ntot/2+1)$.   

In what follows we prove, for any coupling graph with edges $\ES$, the operator inequality\footnote{Note that this doesn't depend on the sign of $J$ because the operator $\Delta H/(4J) = \sum_{(\n,\m)\in \ES}[I-\vec{\sigma}_{\n}\cdot \vec{\sigma}_{\m}]/4 >0$ is independent of $J$. Of course, when $J<0$ we have $\Delta H<0$ and $E_*$ is the maximum energy of $H$ rather than minimum, but the stated inequality holds regardless.}
\begin{align} \label{eq:main_bound}
	\Delta\SvecSq \le  c  \frac{\Delta H}{4J} ,
\end{align}
where $c =  \Ntot (\Ntot-1)^2$.  

For a rectangular lattice $N$ spins wide in each of $D$ dimensions (so that $\Ntot = N^D$) with periodic boundary conditions, we can tighten this to 
\begin{align}\label{eq:periodic}
	c = \frac{3D+1}{48} N^{D+2} + O\left(D N^{D+1}\right)
\end{align}
and in the case of open boundary conditions
\begin{align}\label{eq:open}
	c=  \frac{(3D+1)2^{D}}{12} N^{D+2} + O\left(D 2^D N^{D+1}\right).
\end{align}

These results follow straightforwardly from the following fact:

\begin{lemma}\label{lem:singlet-operator-bound}
For any integer $\m \ge 2$ and any choice of positive coefficients $c_\elle>1$ satisfying $\sum_{\elle=1}^{\m} c_\elle^{-1} = 1$, the singlet projectors for qubits labeled $1,\ldots,k+1$ obey
\begin{align}
	\label{singlet-operator-bound}
	\Psing{1,\m+1} \le \sum_{\elle=1}^{\m} c_\elle \Psing{\elle,\elle+1} 
\end{align}
\end{lemma}

Up to an overall factor of two, Correggi et al.~\cite{correggi2014validity, correggi2015validity} previously proved this in the special case of constant $c_\elle$. (See Lemma~1 of Ref.~\cite{correggi2015validity} and Proposition~5.2 of Ref.~\cite{correggi2015validity}.) They did not try to obtain explicit constants, but they found the same leading $N^{D+2}$ behavior in their case of interest $D=3$ that is seen in \eqref{eq:periodic} and \eqref{eq:open} above. Ref.~\cite{correggi2015validity} also addresses larger spins.

In Sec.~\ref{sec:lemma} we prove Lemma~\ref{lem:singlet-operator-bound}. 
In Sec.~\ref{sec:theorem} we apply it to get the main spin-energy operator inequalities listed above, obtaining constants for the three classes of graphs.  In Sec.~\ref{sec:lit}, we contrast our work with a different spin-energy inequality from B\"arwinkel et al.\ \cite{barwinkel2003improved}, and in Sec.~\ref{sec:discussion} we interpret our result in terms of low-energy magnons.

\section{Singlet-projector inequality for a qubit chain}\label{sec:lemma}

\subsection{Chain of three qubits}\label{sec:three-qubits}

We begin by proving Lemma~\ref{lem:singlet-operator-bound} in the special case of three qubits ($k=2$). In the following subsection we extend this to an arbitrary number of qubits by induction.

Consider just three spins.  Intuitively, if the first and second spins are highly aligned, and if the second and third spins are highly aligned, then the first and third spins should be highly aligned.  More precisely, we want to bound the operator $\vec{\sigma}_1\cdot \vec{\sigma}_3$ from below with $\vec{\sigma}_1\cdot \vec{\sigma}_2$ and $\vec{\sigma}_2\cdot \vec{\sigma}_3$.  Since both $\Delta\SvecSq$ and $\Delta H$ in our eventual inequality are linear combinations of such operators, we seek a linear bound of the form 
\begin{align}
	\vec{\sigma}_1\cdot \vec{\sigma}_3 \ge a\, \vec{\sigma}_1\cdot \vec{\sigma}_2 + b\, \vec{\sigma}_2\cdot \vec{\sigma}_3
\end{align}
for positive numbers\footnote{One might expect we should concentrate on the special case $a=b$ since the first and third spins are on equal footing here, but below it will be helpful to have the more general case in order to correctly combine inequalities across more than three spins. (The case $a=b$ turns out to be sufficient if $N$ is a power of 2 since then the inequalities can be nested rather than chained together.)} $a$ and $b$.
Because $\vec{\sigma}_{\n}\cdot \vec{\sigma}_\m = \Id - 4 \Psing{\n,\m}$ for $\n\neq \m$, an equivalently powerful bound would take the form
\begin{align}
	\Psing{1,3} \le a \Psing{1,2} + b\Psing{2,3}.
\end{align}
One can directly compute the two distinct non-zero eigenvalues (each doubly degenerate) of the 8-dimensional operator $a\Psing{1,2} + b\Psing{2,3} - \Psing{1,3}$ to be
\begin{align}
	\frac{1}{2}\left(-1+a+b\pm\sqrt{1+a+b-ab+a^2+b^2}\right)
\end{align}
which are simultaneously non-negative exactly when $ab \ge a+b$.  
Therefore, we see that the bound 
\begin{align}
	\label{three-qubit-singlet-operator-bound}
	\Psing{1,3} \le \frac{b}{b-1} \Psing{1,2} + b \Psing{2,3}
\end{align}
always holds for any $b>1$.

\subsection{Chain of arbitrary length}\label{sec:arb-qubits}

The previous equation constrains alignment of the spins of next-nearest neighbor spins in terms of the alignment of nearest neighbor spins.  We can chain these inequalities together to prove \eqref{singlet-operator-bound} by induction.

The base case of induction is a chain of $\m+1=3$ qubits connected by $\m=2$ vertices, for which the relevant bound is
\begin{align}
	\Psing{1,3} \le c_1 \Psing{1,2} + c_2 \Psing{2,3}.
\end{align}
whenever $c_1^{-1} + c_2^{-1} = 1$. 
This is just \eqref{three-qubit-singlet-operator-bound}, which we have already proved, with the substitution $b\to c_2$.

Next, we would like to prove \eqref{singlet-operator-bound} holds for arbitrary $\m \ge 3$ and for any choice of $c_\elle > 0$ satisfying $\sum_{\elle=1}^{\m} c_\elle^{-1} = 1$ under the inductive assumption that
\begin{align}
	\label{eq:inductive-assumption}
	\Psing{1,\m} \le \sum_{\elle=1}^{\m-1}c^\prime_\elle \Psing{\elle,\elle+1}
\end{align}
holds for any $c^\prime_\elle>0$ satisfying $\sum_{\elle=1}^{\m-1} (c^\prime_\elle)^{-1} = 1$. 
We can start with \eqref{three-qubit-singlet-operator-bound} (with $b\to c_{\m}$) applied to the three qubits numbered $1$, $k$, and $k+1$ to get
\begin{align}
	\Psing{1,\m+1} &\le \frac{c_{\m}}{c_{\m}-1} \Psing{1,\m} + c_{\m} \Psing{\m,\m+1}\\
	&\le \frac{c_{\m}}{c_{\m}-1} \sum_{\elle=1}^{\m-1} c^\prime_\elle \Psing{\elle,\elle+1} + c_{\m} \Psing{\m,\m+1}\\
	&= \sum_{\elle=1}^{\m}c_\elle \Psing{\elle,\elle+1} ,
\end{align}
where in the second line we have applied the inductive assumption \eqref{eq:inductive-assumption} and in the third line we have made the choice
\begin{align}\label{eq:old-c-l}
	c_\elle^\prime := \frac{c_{\m} -1}{c_{\m}} c_\elle
\end{align}
for $\elle = 1,2,\ldots \m-1$.
We confirm that $\sum_{\elle=1}^{\m} (c_\elle)^{-1} = 1$ implies
\begin{align}
	\sum_{\elle=1}^{\m-1} \frac{1}{c_\elle^\prime} = \frac{c_{\m} }{c_{\m}-1}\left(\sum_{\elle=1}^{\m-1} \frac{1}{c_\elle}\right) = \frac{c_{\m} }{c_{\m}-1}\left(1-\frac{1}{c_{\m}}\right) = 1.
\end{align}
for our choice \eqref{eq:old-c-l}.
Having proved the bound on $\Psing{1,\m+1}$ assuming the bound holds for $\Psing{1,\m}$, and also proved the bound for the base case $\Psing{1,3}$, we conclude that \eqref{singlet-operator-bound} holds for all integers $\m\ge 3$.

\section{Upper bounding the squared spin with the energy}\label{sec:theorem}

Given a graph edge set $\ES$, let $p(\n,\m)$ be a set of neighboring spin pairs defining a shortest path between spins $\n$ and $\m$, so that $|p(\n,\m)|$ is the distance between $\n$ and $\m$ and  $\mathrm{diam}(\ES) := \max_{\n,\m} |p(\n,\m)|$ is the diameter of the graph.  (If there are multiple shortest paths, let $p(\n,\m)$ be one chosen arbitrarily.)

In what follows, we will use our Lemma~\ref{lem:singlet-operator-bound} applied to the singlet operator $\Psing{\n,\m}$ (for various $\n$ and $\m$) with the ($\elle$-independent) choice $c_\elle = |p(\n,\m)|$,
\begin{align}\label{singlet-operator-bound-simplified}
	\Psing{\n,\m} \le |p(\n,\m)| \sum_{(\elle,\elle')\in p(\n,\m)}  \Psing{\elle,\elle'},
\end{align}
which satisfies $\sum_{(\elle,\elle')\in p(\n,\m)} c_\elle^{-1} = |p(\n,\m)|^{-1} \sum_{(\elle,\elle')\in p(\n,\m)} (1) = 1$ trivially.

\subsection{Arbitrary lattice}

First consider an arbitrary graph with edges $\ES$. Starting with \eqref{eq:simple-start-periodic}, the definition of $\Delta\SvecSq$,  we see
\begin{align}
	\Delta\SvecSq 
	&= \sum_{\n} \sum_{\m \neq \n}  \Psing{\n,\m} \\
	\label{eq:uniform-weight-singlet-path-bound}
	&\le \sum_{\n} \sum_{\m \neq \n} |p(\n,\m)| \sum_{(\elle,\elle')\in p(\n,\m)}  \Psing{\elle,\elle'} \\
	&\le \mathrm{diam}(\ES) \sum_{\n} \sum_{\m \neq \n}  \sum_{(\elle,\elle')\in p(\n,\m)} \Psing{\elle,\elle'} \\ 
	&\le \mathrm{diam}(\ES) \sum_{\n} \sum_{\m \neq \n}  \frac{\Delta H}{4J} \\ 
	&\le \mathrm{diam}(\ES) \Ntot(\Ntot-1)  \frac{\Delta H}{4J} \\
	&\le \Ntot (\Ntot-1)^2  \frac{\Delta H}{4J} \\
	&\le \left[\Ntot^3 +  O(\Ntot^2)\right] \frac{\Delta H}{4J} 
\end{align}
where we have used qubit-chain inequality \eqref{singlet-operator-bound}, 
the fact that $|p(\n,\m)| \le \mathrm{diam}(\ES)$, 
the fact that the singlet-projector for each edge on a non-intersecting path through $\ES$ appears at most once in the sum $\Delta H/4J = \sum_{(\elle,\elle')\in \ES} \Psing{\elle,\elle'} \ge \sum_{(\elle,\elle')\in p(\n,\m)} \Psing{\elle,\elle'}$, 
and the fact that $\mathrm{diam}(\ES) \le \Ntot -1$.  

\subsection{Rectangular lattice with periodic boundary conditions}

This bound for an arbitrary lattice is rather loose in more than one dimension since the fraction of all singlet projectors in $\Delta H$ that appear in any given shortest path $p(\n,\m)$ becomes very small for large $\Ntot$ in most types of graphs.  Therefore let us specialize to a rectangular lattice $N$ spins wide in each of $D$ dimensions, so that $\Ntot = N^D$.  We use the vector notation $\vec \n=(\n_1,\ldots, \n_D)$ with $|\, \vec{\n}\,|_1 = \sum_{d=1}^D |\n_d|$ and use $\hat{d} = (0,\ldots, 0, 1, 0, \dots 0)$ to denote the unit vector in the direction of the $d$-th dimension.  We define the shorthand $P(\vec \n; \vec{\Delta \m}) :=  \Psing{\vec \n,\vec \n+\vec{\Delta \m}}$, using wrap-around indexing for periodic boundary conditions when any of the integer components of $\vec \n$ are outside the range $1$ to $N$.   
Then for periodic boundary conditions we start from \eqref{eq:uniform-weight-singlet-path-bound} and get
\begin{align}
	\Delta\SvecSq 
	&\le \sum_{\vec \n}\sum_{\vec \m \neq \vec \n}  |p(\vec \n,\vec \m)| \sum_{(\vec \elle,\vec \elle')\in p(\vec \n,\vec \m)} \Psing{\vec \elle,\vec \elle'} \\
	\label{eq:square-path}
	&= \sum_{\vec \n}\sum_{\vec{\Delta \m}}  |\vec{\Delta \m}|_1 \left[\sum_{d=1}^D \sum_{\w= 0}^{\Delta \m_d - 1} P\left(\vec{\n} + \sum_{d'=1}^{d-1} \Delta \m_{d'} \hat{d'} + \w \hat{d}; \hat{d}\right)\right] \\
	&= \sum_{d=1}^D \sum_{\vec{\Delta \m}}  |\vec{\Delta \m}|_1 \sum_{\w = 0}^{\Delta \m_d - 1} \sum_{\vec \n} P\left(\vec{\n} + \sum_{d'=1}^{d-1} \Delta \m_{d'} \hat{d'} + \w \hat{d}; \hat{d}\right)\\
	\label{eq:square-path-simplify}
	&= \sum_{d=1}^D \sum_{\vec{\Delta \m}} |\vec{\Delta \m}|_1 \sum_{\w = 0}^{\Delta \m_d - 1} \sum_{\vec \n} P\left(\vec{\n}; \hat{d}\right)\\
	&=  \sum_{d=1}^D  \sum_{\vec{\Delta \m}}  |\Delta \m_d| |\vec{\Delta \m}|_1  \sum_{\vec \n} P\left(\vec{\n}; \hat{d}\right) \\
	&=  \sum_{d=1}^D \left[ \sum_{\Delta \m_1=-\lfloor(N-1)/2\rfloor}^{\lceil (N-1)/2\rceil} \cdots \sum_{\Delta \m_D=-\lfloor(N-1)/2\rfloor}^{\lceil (N-1)/2\rceil}   |\Delta \m_d|  \sum_{d'=1}^D  |\Delta \m_{d'}|\right] \sum_{\vec \n} P\left(\vec{\n}; \hat{d}\right) \\
	\label{eq:do-delta-H-def}
	&= \sum_{d=1}^D \Bigg[(D-1)N^{D-2} \left(\sum_{\w =-\lfloor(N-1)/2\rfloor}^{\lceil (N-1)/2\rceil} |\w|\right)^2  + N^{D-1} \left(\sum_{\w =-\lfloor(N-1)/2\rfloor}^{\lceil (N-1)/2\rceil} |\w|^2\right) \Bigg] \sum_{\vec \n} P\left(\vec{\n}; \hat{d}\right)\\
	&= N^{D-2} \Bigg[(D-1) \left(\frac{N^2-\delta_N}{4}\right)^2 + N \left(\frac{N(N^2+2-3\delta_N)}{12}\right) \Bigg] \frac{\Delta H}{4J}  \\
	&=  \left[ \frac{3D+1}{48} N^{D+2} + O\left(D N^{D+1}\right)\right]  \frac{\Delta H}{4J}
\end{align}
where $\lfloor\,\cdot\,\rfloor$ and $\lceil\,\cdot\,\rceil$ denote the integer floor and ceiling functions, respectively, and where $\delta_N := N \, \mathrm{mod}\, 2 = \{0\, \mathrm{if}\,N\, \mathrm{even}; 1\, \mathrm{if}\,N\, \mathrm{odd} \}$.
In \eqref{eq:square-path} we changed summation variables from $\vec \m$ to $\vec{\Delta \m} = \vec{\m}-\vec{\n}$ and used $|p(\vec\n,\vec\m)| = |\vec{\m}-\vec{\n}|_1 = |\vec{\Delta \m}|_1$.
In \eqref{eq:square-path-simplify} we used the fact that $\sum_{\vec \n} P(\vec{\n} + \vec{\elle}; \hat{d}) = \sum_{\vec \n} P(\vec{\n}; \hat{d})$ for any fixed $\vec{\elle}$.  
In \eqref{eq:do-delta-H-def}, we completed the integer sums and used $ \sum_{d=1}^D\sum_{\vec \n} P(\vec{\n}; \hat{d})=\Delta H/(4J)$.
Note that in \eqref{eq:square-path}, the expression in square brackets is the sum of all singlet projectors on a shortest path between sites $\vec{\n}$ and $\vec{\m} = \vec{\n}+\vec{\Delta \m}$ that is constructed by starting from $\vec{\n}$ and taking $\Delta \m_1$ nearest-neighbor steps in the first dimension, then $\Delta \m_2$ steps in the second dimension, and so on through the $D$-th dimension until site $\vec{\m}$ is reached.

After making the choices in the first line to (a) bound each $\Psing{\n,\m}$ in $\SvecSq$ with a single shortest path $p(\vec \n,\vec \m)$ and (b) use the uniform path weighting $c^\elle_{\vec \n,\vec \m} = |p(\vec \n,\vec \m)|$, we use no further inequalities.  We think these choices are probably optimal in light of the equivalence (due to translational invariance) of all lattice sites, in which case this seems to be the tightest bound one could prove with a technique based on \eqref{singlet-operator-bound}.

\subsection{Rectangular lattice with open boundary conditions}

The case of open boundary conditions is trickier since we no longer have full translational invariance, and so we give up on possible optimality. We will change convention so that $P(\vec{\elle};\vec{\elle}')$ is taken to vanish when the indices are out of range, which will be convenient for open boundary conditions. We can show the following:
\begin{align}
	\Delta\SvecSq 
	&\le \sum_{\vec \n}\sum_{\vec \m \neq \vec \n}  |p(\vec \n,\vec \m)| \sum_{(\vec \elle,\vec \elle')\in p(\vec \n,\vec \m)} \Psing{\vec \elle,\vec \elle'} \\
	&= \sum_{\n_1=1}^N\sum_{\Delta \m_1 = 1-\n_D}^{N-\n_1}  \cdots \sum_{\n_D=1}^N  \sum_{\Delta \m_D = 1-\n_D}^{N-\n_D} |\vec{\Delta \m}|_1 \sum_{(\vec \elle,\vec \elle')\in p(\vec \n,\vec \m)} \Psing{\vec \elle,\vec \elle'}\\
	&= \sum_{\Delta \m_1 = 1-N}^{N-1} \sum_{\n_1=1-\mathrm{min}(\Delta \m_1,0)}^{N-\mathrm{max}(\Delta \m_1,0)}  \cdots \sum_{\Delta \m_D = 1-N}^{N-1}\sum_{\n_1=1-\mathrm{min}(\Delta \m_D,0)}^{N-\mathrm{max}(\Delta \m_D,0)}  |\vec{\Delta \m}|_1 \sum_{(\vec \elle,\vec \elle')\in p(\vec \n,\vec \m)} \Psing{\vec \elle,\vec \elle'}\\
	&\le \sum_{\Delta \m_1 = 1-N}^{N-1} \cdots \sum_{\Delta \m_D = 1-N}^{N-1}   |\vec{\Delta \m}|_1  \sum_{\vec{\n}} \sum_{(\vec \elle,\vec \elle')\in p(\vec \n,\vec \m)} \Psing{\vec \elle,\vec \elle'}\\
	&=\sum_{\Delta \m_1 = 1-N}^{N-1} \cdots \sum_{\Delta \m_D = 1-N}^{N-1}   |\vec{\Delta \m}|_1 \sum_{\vec{\n}} \sum_{d=1}^D \sum_{\w = 0}^{|\Delta \m_d| - 1}  P\left(\vec{\n} + \sum_{d'=1}^{d-1} \Delta \m_{d'} \hat{d'} + \mathrm{sgn}(\Delta \m_d) \w \hat{d}; \hat{d}\right) \\
	&= \sum_{\Delta \m_1 = 1-N}^{N-1} \cdots \sum_{\Delta \m_D = 1-N}^{N-1}   |\vec{\Delta \m}|_1  \sum_{d=1}^D \sum_{\w = 0}^{|\Delta \m_d| - 1} \sum_{\vec{\n}} P\left(\vec{\n} ; \hat{d}\right) \\
	&= \sum_{\Delta \m_1 = 1-N}^{N-1} \cdots \sum_{\Delta \m_D = 1-N}^{N-1}   |\vec{\Delta \m}|_1  \sum_{d=1}^D |\Delta \m_d| \sum_{\vec{\n}} P\left(\vec{\n} ; \hat{d}\right) \\
	&= \sum_{d=1}^D \left[  \sum_{\Delta \m_1 = 1-N}^{N-1} \cdots \sum_{\Delta \m_D = 1-N}^{N-1}   |\Delta \m_d| \sum_{d'=1}^D  |\Delta \m_{d'}|  \right]\sum_{\vec \n} P\left(\vec{\n}; \hat{d}\right)\\
	&=  \sum_{d=1}^D  \Bigg[(D-1)(2N-1)^{D-2} \left(\sum_{\w =1-N}^{N-1} |\w|\right)^2  + (2N-1)^{D-1} \left(\sum_{\w =1-N}^{N-1}|\w|^2\right) \Bigg] \sum_{\vec \n} P\left(\vec{\n}; \hat{d}\right) \\
	&= (2N-1)^{D-2} \Bigg[(D-1) N^2 (N-1)^2  + (2N-1) \left(\frac{N(N-1)(2N-1)}{3}\right) \Bigg] \frac{\Delta H}{4J}  \\
	&=  \left[ \frac{(3D+1)2^{D}}{12} N^{D+2} + O\left(D 2^D N^{D+1}\right)\right]  \frac{\Delta H}{4J}
\end{align}
In the third line we exchange the order of the $\n_d$ and $\Delta \m_d$ summations, and in the fourth line we expand the range of the summations over the $\n_d$ so that, in the fifth line, we can use $\sum_{\vec \n} P\left(\vec{\n} + \vec{\elle}; \hat{d}\right) = \sum_{\vec \n} P\left(\vec{\n}; \hat{d}\right)$ for any $\vec{\elle}$.

This bound is clearly suboptimal. First, unlike the case of periodic boundary conditions, here we used a second inequality (in the fourth line) as a way of coping with the lack of translational symmetry; the expansion of the summations over the $\n_d$ leads to the factor of $2^D$ in the last line.  The second, closely related issue is that the singlet projectors for nearest-neighbor qubit pairs near the center of the lattice will appear more times in nearest-neighbor paths and hence get more ``weight'' in the sum than those near the periphery of the lattice.  Numerically we have checked that a bound with a tighter constant coefficient can be obtained by choosing $c^\elle_{\vec \n, \vec \m}$ to be smaller for central pairs and larger for peripheral pairs. However, we do not have any significant analytical results along these lines, and numerically we found the improvement in the bound coefficients to be modest in one-dimension. It likely is more important for high dimension.

\section{Comparison to B\"arwinkel et al.}\label{sec:lit}

B\"arwinkel et al.\ \cite{barwinkel2003improved} have proven a class of spin-energy operator inequalities for systems with arbitrary spins (not just qubits) and non-constant couplings.  Here we show that, in the case of spin $1/2$ particles with constant couplings where the above results apply, the inequality is similar to but distinct from that of B\"arwinkel et al. in the sense that one or the other may be stronger depending on the particular configuration.

\newcommand{\J}{ \mathcal{J}}

For the Hamiltonian $H = \sum_{\n,\m} \J_{\n,\m} \vec{s}_{\n}\cdot \vec{s}_\m$ of arbitrary Heisenberg couplings of strength $\J_{\n,\m} = \J_{\m,\n}$ between spins $\vec{s}_{\n}$ of magnitude $s$ labeled by $\n$, B\"arwinkel et al.\ prove an energy constraint under a condition they call 
``$\en$-homogeneity'' 
\cite{barwinkel2003improved}. We are concerned with situation where all matrix elements of $\J$ are 0 or 1, for which only the case $\en=1$ is applicable.  This was called ``weak homogeneity'' in earlier work \cite{schmidt2001bounding}, and is the condition that $j:= \sum_{\n} \J_{\n,\m} $ is a constant independent of $\m$.  This can be satisfied without changing the Hamiltonian by re-defining the on-diagonal components of $\J$ as $\J^{\mathrm{new}}_{\elle,\elle}:= \Ntot^{-1} \sum_{\n,\m} \J^{\mathrm{old}}_{\n,\m} - \sum_{\n\neq \elle} \J^{\mathrm{old}}_{\n,\elle}$.  Weak homogeneity is thus only a gauge condition for fixing a choice of matrix $\J$ given a Hamiltonian.

Assuming the gauge of $\J$ is set thus, the bound is\footnote{Their bound is a strengthening of Ref.~\cite{schmidt2001bounding}.  For related work, see Refs.~\cite{schnack2001independent,barwinkel2002energy,schmidt2002linear}.} 
\begin{align}
	\label{barwinkel-bound}
	H \ge \frac{j-j_{\mathrm{min}}}{\Ntot}\vec{S}^2 + N_{\mathrm{tot}}  j_{\mathrm{min}} s(s+1) + (N_{\mathrm{tot}} -1)(j_2-j_{\mathrm{min}})s
\end{align}
where $j = \sum_{\n} \J_{\n,\m} $ is the eigenvalue of the matrix $\J$ associated with the ``all-ones'' vector $(1,1,\ldots,1)$, $j_{\mathrm{min}}$ is the smallest eigenvalue in the subspace perpendicular to $(1,1,\ldots,1)$, and $j_2$ is the next smallest eigenvalue.

For qubits, $s=1/2$ and $ \vec{s}_{\n}\cdot \vec{s}_\m = \vec{\sigma}_{\n}\cdot \vec{\sigma}_\m/4$. In the special case of a $D$-dimensional lattice of $N_{\mathrm{tot}} = N^D$ qubits with periodic boundary conditions and constant nearest-neighbor couplings $\J_{\vec{\n},\vec{\m}} = -2J \delta_{|\vec{\n}-\vec{\m}|_1 = 1}$, we get
\begin{align}
	j &= -4 J D,\\
	j_{\mathrm{min}} &= j_2 =  -4 J [D-1+\cos(2\pi/N)] .
\end{align}
With this we can losslessly covert \eqref{barwinkel-bound} to a bound on $\Delta\SvecSq:= \SvecSqMax-\SvecSq$ in terms of $\Delta H = H - E_*$ (where here $E_* = -2 J D  N_{\mathrm{tot}} = -2 J D N^D$):
\begin{align}
	\Delta\SvecSq &\le \left[\frac{N^D}{1-\cos(2\pi/N)}\right] \frac{\Delta H}{4J} + \left(\frac{D}{2[1-\cos(2\pi/N)]}-\frac{1}{2}\right)N^{2D} + \frac{1}{2} N^D\\
	&\le \left[\frac{1}{2\pi^2} N^{D+2} + O(N^D)\right] \frac{\Delta H}{4J} + \frac{D}{4\pi^2} N^{2D+2} + O(N^{2D})
\end{align}
where we have written down just the leading-order behavior in $N$ for simplicity. 

The term that goes like $N^{D+2}\Delta H$ is similar to our bound \eqref{eq:periodic}, but the term that goes like $N^{2D+2}$ is not present.  Therefore, when $\Delta H / 4J \ll N^D$, our bound is tighter. On the other hand, the Hamiltonian is bounded by  $\Delta H/4J = \sum_{(\n,\m)\in \ES} = \Psing{\n,\m} \le D\Ntot I = D N^D I$ (because $D N^D $ is the number of bonds in the lattice), and when it is comparable to this constraint the bound of  B\"arwinkel et al.\ has a slightly better coefficient, especially for large dimension: $(\Delta H/4J )N^{D+2}/(2\pi^2) + DN^{2D+2}/(4\pi^2) \sim (3DN^{2D+2}/4\pi^2) < (3D+1)DN^{2D+2}/48 \sim (\Delta H/4J )(3D+1)N^{D+2}/48$.

\section{Discussion}\label{sec:discussion}
For the ferromagnetic Heisenberg model on a $d$-dimensional lattice, we can loosely interpret the bound in~\eqref{eq:main_bound},~\eqref{eq:periodic} as a statement about magnon excitations.  First we briefly recall the notion of magnons, the bosonic excitations that constitute spin waves. The ferromagnetic groundspace is degenerate; we focus on a single groundstate $|\psi_0\rangle = |\downarrow\rangle^{\otimes \Ntot}$.  Denoting positions on the lattice by $\vec{x} \in \{1,\ldots,N\}^D$, define the spin ``creation'' operator 
\begin{align}
S_{\vec{k}}^+ = \frac{1}{\sqrt{\Ntot}} \sum_{\vec{x}} e^{i \vec{k} \cdot \vec{x}} \sigma^x_{\vec{x}}
\end{align}
parameterized by lattice momentum $\vec{k}$ with $k_i=\frac{2 \pi m}{N}$ for $m \in \mathbb{Z}_N$.  Then the states $S^+_{\vec{k}} |\psi_0\rangle$ are eigenstates, referred to as single-magnon excitations, with energy approximately $2J\vec{k}^2$ above the groundstate for small $\vec{k}$.   
More generally, states $S^+_{{\vec{k}}_1}\ldots S^+_{{\vec{k}}_m}  |\psi_0 \rangle$ for momenta $\vec{k}_1,\ldots,\vec{k}_m$ may be loosely interpreted as $m$-magnon states.  (For $m>1$, these are not precise eigenstates, essentially due to interactions between magnons.) At long wavelength and low energy, the magnons are approximately non-interacting, and the energy of the above multi-magnon state is roughly the sum of the energies associated with the individual magnons. The study of the Heisenberg model under these approximations is known as “spin wave theory” \cite{dyson1956general,correggi2014validity}.  

Returning to the spin-energy inequality, for a $D$-dimensional rectangular lattice of $N$ spins with $\Ntot =N^D$, we rearrange \eqref{eq:main_bound} and \eqref{eq:periodic} as
\begin{align} \label{eq:magnon_bound}
\left( \frac{\Delta\SvecSq}{\Ntot} \right) \left( \frac{J}{N^2} \right)  \le c_0 \Delta H
\end{align}
for some constant $c_0$ independent of $N$.  
The ground space is an eigenspace of $\SvecSq$ with maximum spin $\SvecSq=\SvecSqMax=\frac{\Ntot}{2}(\frac{\Ntot}{2}+1)$.  The subspace of single-magnon states described above have the next-largest $\SvecSq $ eigenvalue of $\frac{\Ntot-1}{2}(\frac{\Ntot-1}{2}+1)$, corresponding to $\Delta \SvecSq = \frac{1}{2}\Ntot +\frac{1}{4}$.  More generally, for states with $m$ magnons and $m \ll N$ (so that the spin-wave approximation holds), one has $\Delta \SvecSq \propto m \Ntot$.  Then the first factor on the left-hand side of \eqref{eq:magnon_bound} is proportional to the number of magnons, or the number of spin flips with respect to the vacuum state.
Meanwhile, we can also interpret the second factor of \eqref{eq:magnon_bound} in terms of magnons:  the lowest energy magnons have wavelength $N$ and energy like $\frac{J}{N^2}$.  So \eqref{eq:magnon_bound} roughly states: the energy $\Delta H$ is lower bounded by the number of magnons times the smallest possible energy per magnon. From this perspective, the bound seems straightforward, but of course the above picture is only heuristic, while the bound provides a precise operator inequality.

The magnon analysis also implies the bound depends optimally on $N$.  (See also the discussion of optimality in Proposition~5.2 of Ref.~\cite{correggi2015validity}.) In particular, the leading term $N^{D+2}$ in \eqref{eq:periodic} must have exponent at least $D+2$, due to the existence of the single-magnon eigenstates discussed above. 

\section{Acknowledgments}
D.R.\ acknowledges support
from NTT (Grant AGMT DTD 9/24/20).

\bibliographystyle{abbrv} 
\bibliography{fph}

\end{document}